\documentclass[aps,twocolumn,pra,superscriptaddress,showpacs,showkeys,floatfix]{revtex4-1}
\usepackage{amsmath,amsfonts,amssymb,graphicx}
\usepackage{color}
\usepackage{graphicx}
\usepackage{subfigure}
\usepackage{epsfig}
\usepackage{amsmath}
\usepackage[normalem]{ulem} 
\usepackage{braket}

\newcommand{\Tr}[1]{\mathrm{Tr} #1}

\allowdisplaybreaks

\begin{document}

\title{Experimental detection of quantum channel capacities}

\author{\'Alvaro Cuevas}
\affiliation{:Quantum Optics Group, Dipartimento di Fisica, Universit\`a di Roma La Sapienza,\\
Piazzale Aldo Moro 5, I-00185 Roma, Italy.}
\author{Massimiliano Proietti}
\affiliation{:Edinburgh Mostly Quantum Lab, School of Engineering and Physical Sciences, Heriot-Watt University, David Brewster Building, EH14 4AS Edinburgh, United Kingdom}
\affiliation{:Quantum Optics Group, Dipartimento di Fisica, Universit\`a di Roma La Sapienza,\\
Piazzale Aldo Moro 5, I-00185 Roma, Italy.}
\author{Mario Arnolfo Ciampini}
\affiliation{:Quantum Optics Group, Dipartimento di Fisica, Universit\`a di Roma La Sapienza,\\
Piazzale Aldo Moro 5, I-00185 Roma, Italy.}
\author{Stefano Duranti}
\affiliation{:Quantum Optics Group, Dipartimento di Fisica, Universit\`a di Roma La Sapienza,\\
Piazzale Aldo Moro 5, I-00185 Roma, Italy.}
\affiliation{:Dipartimento di Fisica e Geologia, Universit\`a degli Studi di Perugia,\\
Via Pascoli snc, I-06123 Perugia, Italy.}
\author{Paolo Mataloni}
\affiliation{:Quantum Optics Group, Dipartimento di Fisica, Universit\`a di Roma La Sapienza,\\
Piazzale Aldo Moro 5, I-00185 Roma, Italy.}
\author{Massimiliano F. Sacchi}
\affiliation{:Istituto di Fotonica e Nanotecnologie - Consiglio Nazionale delle Ricerche, 
\\Piazza Leonardo da Vinci 32, I-20133, Milano, Italy.}
\author{Chiara Macchiavello}
\affiliation{:Dipartimento di Fisica, Universit\`a di Pavia, and INFN - Sezione di Pavia\\ 
Via A. Bassi 6,  I-27100 Pavia, Italy}
	
\date{\today}

\begin{abstract} 

We present an efficient experimental procedure that certifies non vanishing 
quantum capacities for qubit noisy channels. Our method is based on the use 
of a fixed bipartite entangled state, where the system qubit is sent to the 
channel input. A particular set of local measurements is performed at the channel 
output and the ancilla qubit mode, obtaining lower bounds to the quantum capacities for any unknown 
channel with no need of a quantum process tomography. The entangled qubits 
have a Bell state configuration and are encoded in photon polarization. 
The lower bounds are found by estimating the Shannon and von Neumann 
entropies at the output using an optimized basis, whose 
statistics is obtained by measuring only the three observables 
$\sigma_{x}\otimes\sigma_{x}$, $\sigma_{y}\otimes\sigma_{y}$ and $\sigma_{z}\otimes\sigma_{z}$. 

\end{abstract}

\maketitle
\textbf{Introduction:} Any communication channel is unavoidably affected by noise that limits its 
ability to transmit information, quantified in terms of channel capacity. 
When the use of the channel aims to convey quantum information, its efficiency 
is evaluated in terms of the quantum capacity, which is the maximum number of 
qubits that can be reliably transmitted per channel use 
\cite{lloyd,barnum,devetak,hay}, and represents a central quantitative notion
in quantum communications. In general the computation of the quantum capacity is a hard task
since it requires a regularisation procedure over an infinite number
of channel uses, and it is therefore by itself not directly accessible
experimentally. Its analytical value is known mainly for some channels
that have the property of degradability \cite{deveshor,ydh,rusk}, since
regularisation is not needed in this case.

In this Letter we address the issue of experimental detection of
Quantum Channel Capacities.  For a generic unknown channel the quantum
capacity can be in principle estimated via quantum process tomography 
\cite{nielsen97,pcz,mls,dlp,alt,cnot,vibr,
ion,mohseni,irene, atom}, 
which provides a complete reconstruction of 
the channel, and therefore leads to an evaluation of all its
communication properties.  This, however, is a demanding procedure in
terms of the number of required different measurement settings, since
it scales as $d^4$ for a finite $d$-dimensional quantum system. Moreover, being an indirect method,
it also has the drawback of involving larger errors due to
error propagation.

Here, we are not interested in reconstructing the complete form of the
noise affecting the channel, but only in detecting its quantum
capacity, which is a very specific feature for which we developed a
novel and less demanding procedure in terms of resources
(measurements) involved. This is pursued in the same spirit as it is
done, for example, in entanglement detection for composite systems
\cite{ent-wit}, in parameter estimation procedures \cite{parest}, and
in the detection of specific properties of quantum channels, such as
being entanglement-breaking \cite{qchanndet} or non-Markovian
\cite{nomadet}.

In this Letter we report the first experiment where a lower bound to
the quantum channel capacity is directly accessed by means of a number
of local measurements that scales as $d^2$, hence more favourably w.r.t. 
process tomography that scales as $d^4$. The experiment is based on
a recently proposed theoretical method \cite{ms16} that can be applied
to generally unknown noisy channels in arbitrary finite dimension, and
has been proved to be very efficient for many examples of qubit
channels \cite{ms-corr}.

The method is suited for any kind of physical system
available for quantum communication, and the experiment we present
here is based on a quantum optical implementation for various forms of
noisy single-qubit channels.\\

\textbf{Lower Bound on Quantum Channel Capacity:} 
The quantum capacity $Q$ of a noisy channel 
${\cal E}$, measured in qubits per channel use, is defined as 
\cite{lloyd,barnum,devetak,hay}
\begin{eqnarray} 
Q=\lim _{N\to \infty}\frac{Q_N}{N}\;,\label{qn} 
\end{eqnarray} 
where $N$ is the number of channel uses, $Q_N = \max_{\rho } I_c (\rho , 
{\cal E}_N)$, with ${\cal E}_N = {\cal E}^{\otimes N}$, and $I_c(\rho , 
{\cal E}_N)$ denotes the coherent information \cite{schumachernielsen}
\begin{equation}
I_c(\rho , {\cal E}_N) = S[{\cal E}_N (\rho )] - S_e (\rho, {\cal E}_N)\;.
\label{ic}
\end{equation}
In the above equation $S(\rho )=-\Tr [\rho \log _2 \rho ]$ is the von Neumann 
entropy and $S_e (\rho, {\cal E})$ represents the entropy exchange 
\cite{schumacher}, i.e. $S_e (\rho, {\cal E})= S[({\cal I}_R \otimes {\cal E})(|\Psi _\rho \rangle \langle \Psi _\rho |)] $, 
where $|\Psi _\rho \rangle $ denotes 
any purification of $\rho $ by means of an ancilla reference quantum system 
$A$, namely $\rho =\Tr _A [|\Psi _\rho \rangle \langle \Psi _\rho|]$.

The following chain of bounds holds
\begin{eqnarray}
Q \geq Q_1 \geq I_c(\rho , {\cal E}_1)\geq Q_{DET}
\;,\label{qvec}
\end{eqnarray}
where the first two inequalities come directly from the above definitions, 
while the last one was proved in \cite{ms16}, with 
\begin{eqnarray}
Q_{DET}=S\left [{\cal E} (\rho )\right ]-H(\vec p)\;.\label{QDET} 
\end{eqnarray}
Here ${\cal E} (\rho )$ is the output state for a single 
use of the channel and $H(\vec p)$ denotes the Shannon entropy for the vector 
of the probabilities $\{p_i\}$ corresponding to a measurement on orthogonal 
projectors $\{ |\Phi _i \rangle \}$ in the tensor product of the ancilla 
and the system Hilbert spaces: 
\begin{eqnarray}
p_i = \Tr [({\cal I}_A \otimes {\cal
E})(|\Psi _\rho \rangle \langle \Psi _\rho |) |\Phi _i \rangle\langle\Phi _i|] 
\;.
\label{pimeas}
\end{eqnarray}
The procedure to detect the lower bound $Q_{DET}$ is the following: 
1) prepare a bipartite pure state $|\Psi _\rho \rangle $; 2) send it through 
the channel ${\cal I} _A\otimes {\cal E}$, where the unknown channel 
${\cal E}$ acts on one of the two subsystems; 3) measure suitable local 
observables on the joint output state in order to estimate 
$S\left [{\cal E} (\rho )\right ]$ and $\vec p$, and to compute $Q_{DET}$. 
After the measurements have been 
performed, the detected bound  $Q_{DET}$ can then be optimized over all 
probability distributions that can be obtained from the used measurement settings. 
This last step is achieved by performing an ordinary classical 
processing of the measurement outcomes.
\begin{figure}[h!]
	\centering
		\includegraphics[width=0.45\textwidth, height=0.195\textwidth]{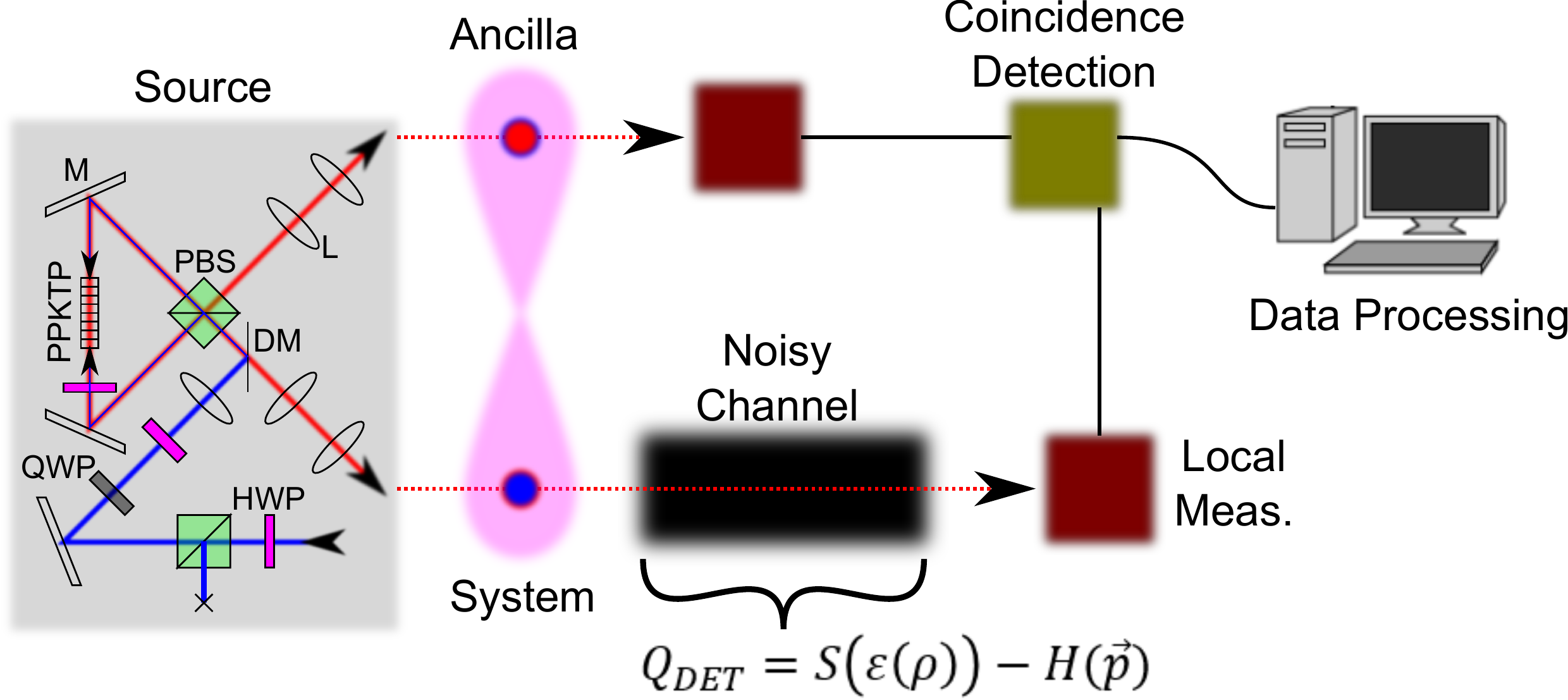}
	\caption{\textbf{Experimental Setup.} A Sagnac interferometric
	source of polarization entangled photons sends the S-qubit
	through the noisy channel, while the A-qubit remains
	untouched. Both qubits are measured in a joint photon counter system. Here PPKTP is a periodically poled
	potassium titanyl phosphate non-linear crystal, PBS a
	polarizing beam splitter, M a mirror, L a converging lens,
	HWP a half-wave plate, QWP a quarter-wave
	plate, and DM a dichroic mirror.}
	\label{fig:setup}
\end{figure}
We now specify our scenario to qubit channels
($d=2$), where the protocol only requires
$d^{2}-1=3$ observables, in our case $\sigma_{x}\otimes\sigma_{x}$,
$\sigma_{y}\otimes\sigma_{y}$, and $\sigma_{z}\otimes\sigma_{z}$ on both 
the ancilla and system qubit.  We also consider a maximally
entangled input state $|\Phi ^+ \rangle ={1}/{\sqrt 2}(|00 \rangle +
|11 \rangle )$. The schematic representation of the procedure is
shown in Fig. \ref{fig:setup}.
The above observables allow one to measure $\{\sigma_i\}$  on the 
system alone by ignoring the statistics of the measurement results on the 
ancilla. Since this set is tomographically complete, 
the system output state ${\cal E} (\rho )$ can be reconstructed, and 
therefore the term $S\left [{\cal E} (\rho )\right ]$ in Eq. (\ref{QDET}) can 
be exactly estimated. 
Moreover, the measurement settings $\{\sigma _x \otimes \sigma
_x, \sigma _y \otimes \sigma _y,\sigma _z \otimes \sigma _z \}$ allow
us to estimate the vector $\vec p$ pertaining to the projectors onto
the following inequivalent bases \cite{ms16} 
\begin{eqnarray}
B_{1}=&\{\ket{B_{1,1}},\ket{B_{1,2}},\ket{B_{1,3}},\ket{B_{1,4}}\}\nonumber\\
=&\{a|\Phi^{+}\rangle+b|\Phi^{-}\rangle,-b|\Phi^{+}\rangle+a|\Phi^{-}\rangle,\label{b1}\\
&c|\Psi^{+}\rangle+d|\Psi^{-}\rangle,-d|\Psi^{+}\rangle+c|\Psi^{-}\rangle\};\nonumber\\
B_{2}=& \{\ket{B_{2,1}},\ket{B_{2,2}},\ket{B_{2,3}},\ket{B_{2,4}}\},\nonumber\\
=&\{a|\Phi^{+}\rangle+b|\Psi^{+}\rangle,-b|\Phi^{+}\rangle+a|\Psi^{+}\rangle,\label{b2}\\
&c|\Phi^{-}\rangle+d|\Psi^{-}\rangle,-d|\Phi^{-}\rangle+c|\Psi^{-}\rangle\};\nonumber\\
B_{3}=& \{\ket{B_{3,1}},\ket{B_{3,2}},\ket{B_{3,3}},\ket{B_{3,4}}\}\nonumber\\
=&\{a|\Phi^{+}\rangle+ib|\Psi^{-}\rangle,ib|\Phi^{+}\rangle+a|\Psi^{-}\rangle,\label{b3}\\
&c|\Phi^{-}\rangle+id|\Psi^{+}\rangle,id|\Phi^{-}\rangle+c|\Psi^{+}\rangle\};\nonumber 
\end{eqnarray} 
where $| \Phi ^\pm \rangle ={1}/{\sqrt 2}(|00
\rangle \pm |11 \rangle )$ and $|\Psi ^\pm \rangle ={1}/{\sqrt 2}(|01
\rangle \pm |10 \rangle )$ denote the Bell states, 
and $a,b,c,d$ are real numbers, such that $a^2+b^2=c^2+d^2=1$. 	

The evaluation of the Shannon entropy $H(\vec{p})$ in the $B_{i}$ basis is 
obtained from its definition
\begin{equation}
H(\vec{p_{i}})=-\sum_{j} p_{i,j}\log_{2}p_{i,j}\,,
\end{equation}
where $\vec{p_{i}}=\{p_{i,j}\}$ is the probability vector associated to $({\cal I} _A \otimes {\cal E})(|\Phi ^+ \rangle \langle \Phi ^+ |)$,  
described in Eq. (\ref{pimeas}). 

The probability vectors can be obtained by measuring the expression
$p_{i,j}=\braket{\Pi_{i,j}}$ as described in the Supplemental
Material \cite{SM}, with $\Pi_{i,j}=\ket{B_{i,j}}\bra{B_{i,j}}$ the
projector on the specific basis element $\ket{B_{i,j}}$. All
expectation values $\braket{.}$ are evaluated for the joint output
state $({\cal I} _A \otimes {\cal E})(|\Phi ^+ \rangle \langle \Phi ^+
|)$ and, using the normalization constraints among $a,b,c,$ and $d$, it
can be demonstrated that all probabilities $p_{i,j}$ depend only on
two real parameters, $b$ and $d$. After collecting the measurement
outcomes, the bound on $Q$ is then maximized over the three bases
$B_1,B_2,B_3$, and by varying $b$ and $d$:
\begin{eqnarray}
Q_{DET}&=&\max _{i=1,2,3}\max _{b,d}Q_{DET}(B_i,b,d) \nonumber \\
&=&S[\mathcal{E}(\rho)]-\min _{i=1,2,3}\min _{b,d}
H[\vec{p}(B_i,b,d)]\;
\end{eqnarray}
This last step is performed by classical processing of 
the measurement outcomes, from which  the set of expectation values
$\{\braket{\mathbb{I}\otimes\sigma_{\alpha}}$, $\braket{\sigma_{\alpha}
\otimes\mathbb{I}}$, $\braket{\sigma_{\alpha}\otimes\sigma_{\alpha}}$,
$\alpha=x,y,x$\} is obtained. 
Differently from a complete process tomography, we 
remark that we do not need to measure the six 
observables of the kind $\sigma _\alpha \otimes \sigma _\beta $ with
$\alpha \neq \beta$ and, moreover, the bound is directly obtained from 
the measured expectations, 
without need of linear inversion and/or maximum likelihood technique. 
Let us also 
notice that the use of an entangled input state in our procedure is not mandatory. 
In fact, the ancilla is locally measured and this is equivalent to herald a single-photon 
state at the channel \cite{nota}. 
\\

\textbf{Experimental Procedure:} We implemented the method using a SPDC source of high purity polarization entangled photons \cite{source} schematically represented in Fig. \ref{fig:setup}, where the qubits were encoded as $\{\ket{0}\equiv\ket{H},\ket{1}\equiv\ket{V}\}$, with $|H\rangle$ ($|V\rangle$) the horizontal (vertical) polarization.

From a continuous wave laser pumping at $405nm$ and bandwidth $<0.01pm$ we generate down-converted pairs of single photons at $810nm$ and bandwidth $\approx0.42nm$, hence entangled in Bell states with a measured fidelity of $F_{exp}=0.979\pm 0.011$,
calculated as reported in \cite{SM} and using standard tomography analysis \cite{tomography}. The entanglement degree of the generated photons corresponds to an average concurrence value of $C_{exp}=0.973\pm0.004$ \cite{liquidC}. 

The input state is ideally the maximally entangled state
$\ket{\Phi^{+}}$. Due to experimental imperfections, the
resulting state can be described by a Werner state
$\rho_W=\frac{4F-1}{3}\ket{\Phi^{+}}
\bra{\Phi^{+}}+\frac{1-F}{3}I\otimes I$, where $F$ is the fidelity
with respect to $\ket{\Phi^{+}}$. The derivation of new detectable bounds suited for bipartite input
states which are affected by isotropic noise 
is reported in \cite{SM}. Such bounds will be used to
compare our experimental results with the theoretical predictions.

The procedure was tested for the following types of noise: Amplitude Damping Channel (ADC), Phase Damping Channel (PDC), Depolarizing Channel (DC), and Pauli Channel (PC) \cite{nielsen-chuang,desurvire}. As remarked above, we employ 3 versus 9 measurements on the output state, as in a usual process tomography.

The polarization based measurement is performed via a Quarter Wave Plate (QWP), a Half Wave Plate (HWP), and a
Polarization Beam Splitter (PBS) located in both the reference (ancilla-A) and principal system (system-S) paths. 

The expectation values $\braket{\sigma_{\alpha }\otimes\sigma_{\alpha }}$ (with $\alpha =x,y,z$) were obtained by
taking every matrix element of $\sigma_{\alpha}\otimes\sigma_{\alpha}$ as a
particular projection of the transmitted state (\cite{SM}), which
was measured by integrating photon coincidences during a time
interval of $5$sec. Since only a set of $12$ measurements of the
state are needed (4 by each of the 3 observables), the $Q_{DET}$ can be obtained in only $60$sec. See \cite{SM} for information about actual photon count rates and detection efficiencies.
\begin{figure}[h!]
  	a){\includegraphics[width=0.2\textwidth,height=0.135\textheight]{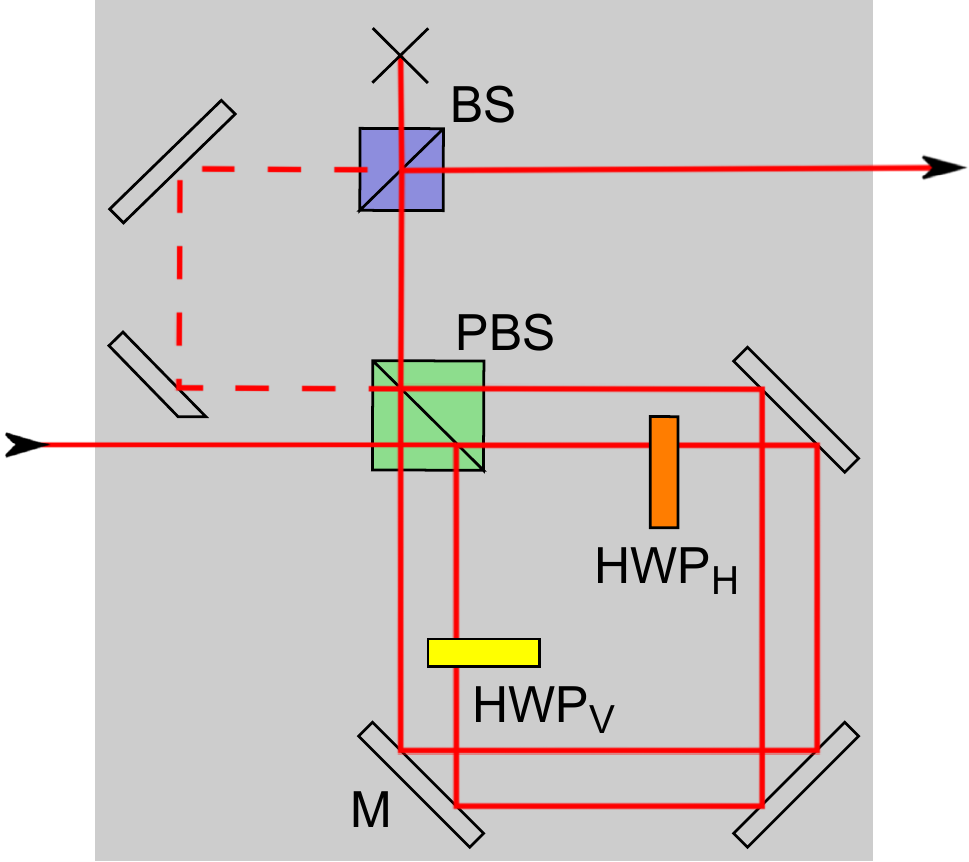}}
		b){\includegraphics[width=0.2\textwidth,height=0.135\textheight]{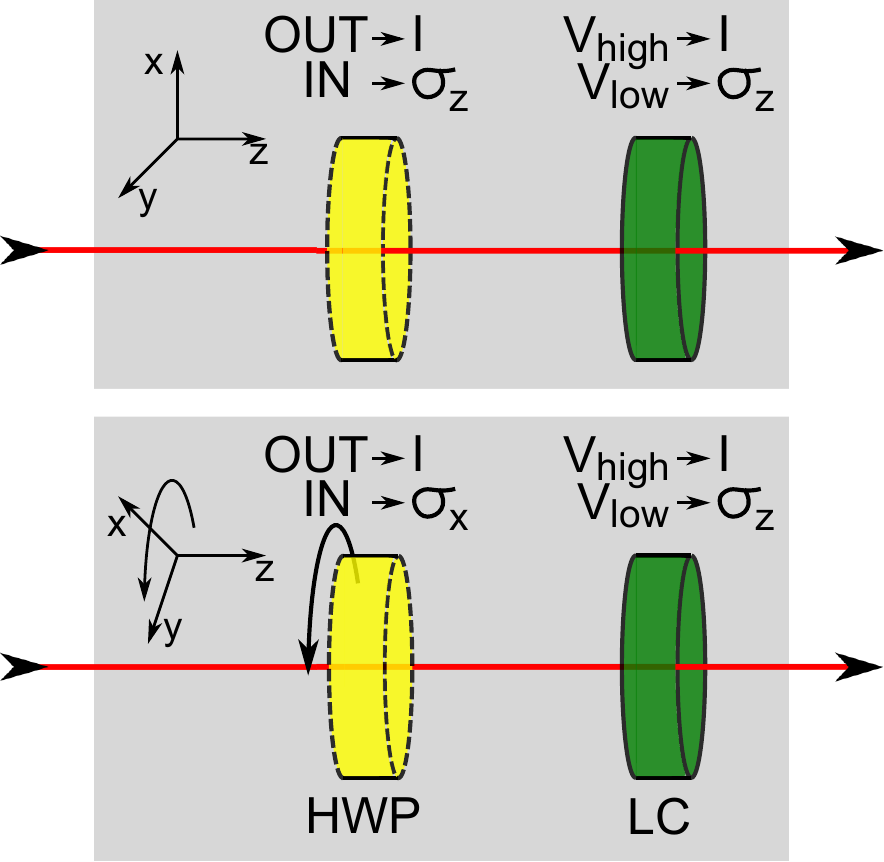}}
	\caption{\textbf{Noisy Channels. a)} Experimental scheme for a 
polarization ADC. Given a single input qubit 
$\ket{\psi}=\alpha\ket{H}+\beta\ket{V}$, it uses two HWPs inside a 
SI to apply a rotation on $\ket{V}$, while $\ket{H}$ 
remains unrotated, achieving $K_{0}$ and $K_{1}$. The dashed line inside the MZI represents the rotated $\sqrt{\gamma}$ 
portion of $\ket{V}$. \textbf{b)-top} Experimental scheme for a PDC. It applies $\mathbb{I}$ and $\sigma_{z}$ over $\ket{\psi}$ by using 
an unrotated HWP and a variable voltage LC. \textbf{b)-bottom:} 
Experimental scheme for the PC and DC. They apply 
$\mathbb{I}$, $\sigma_{x}$, $\sigma_{y}$ and $\sigma_{z}$ over $\ket{\psi}$ 
by using a $45^{\circ}$-rotated HWP and the same LC.}
	\label{fig:channels}
\end{figure}

\begin{figure*}[h!t!b!p!]
a){\includegraphics[width=0.42\textwidth, height=0.26\textwidth]{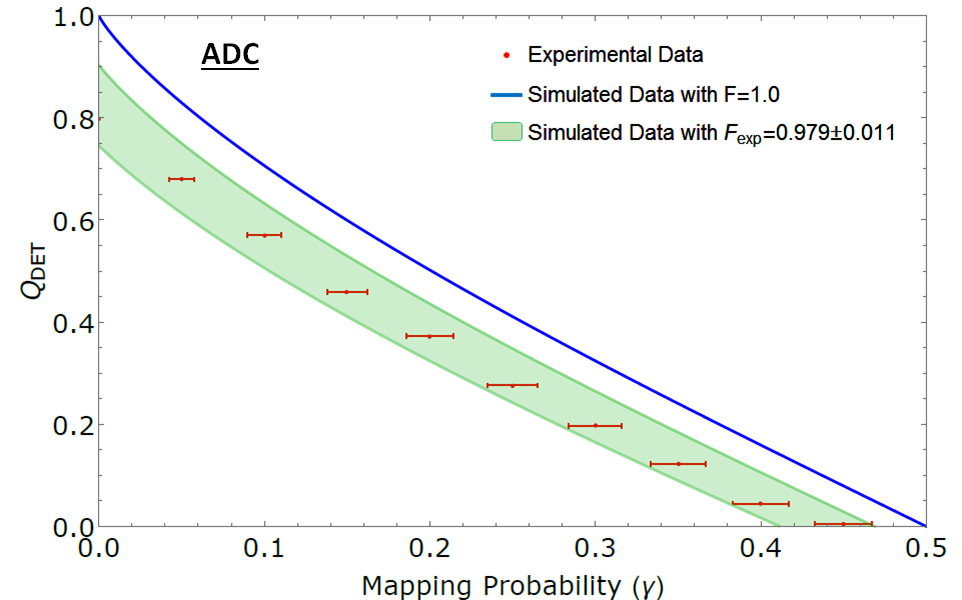}}
b){\includegraphics[width=0.42\textwidth, height=0.26\textwidth]{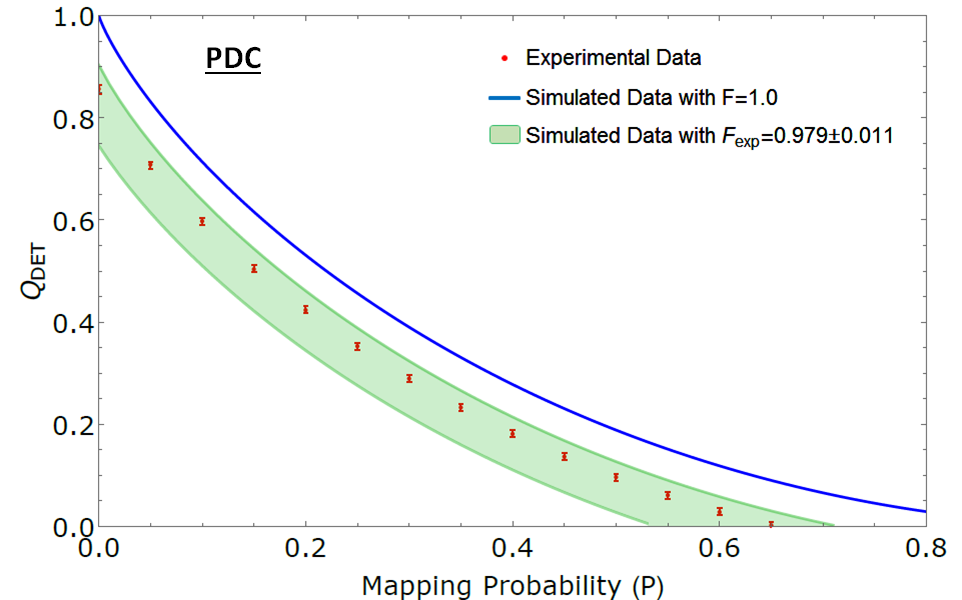}}\\
c){\includegraphics[width=0.42\textwidth, height=0.26\textwidth]{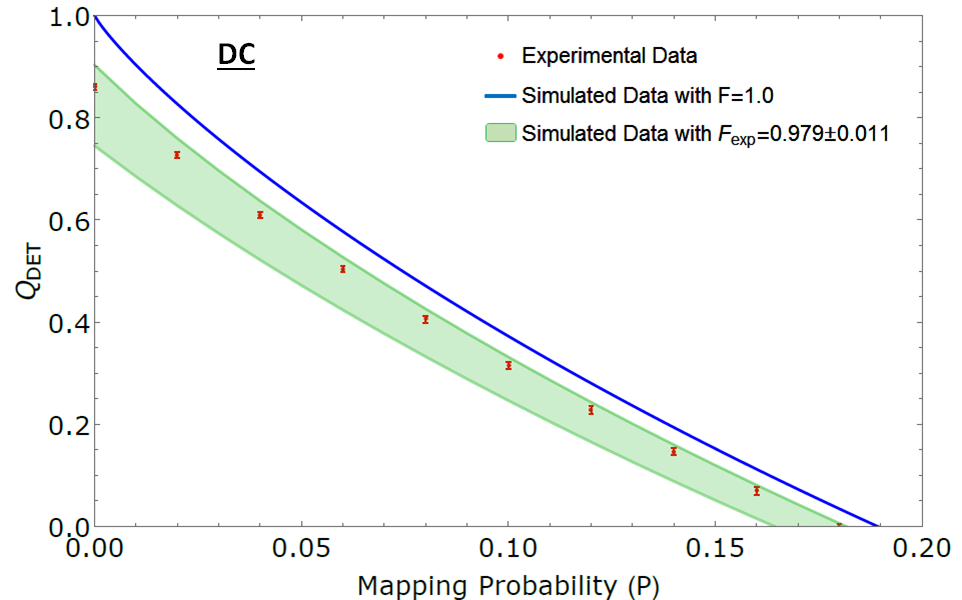}}
d){\includegraphics[width=0.42\textwidth, height=0.26\textwidth]{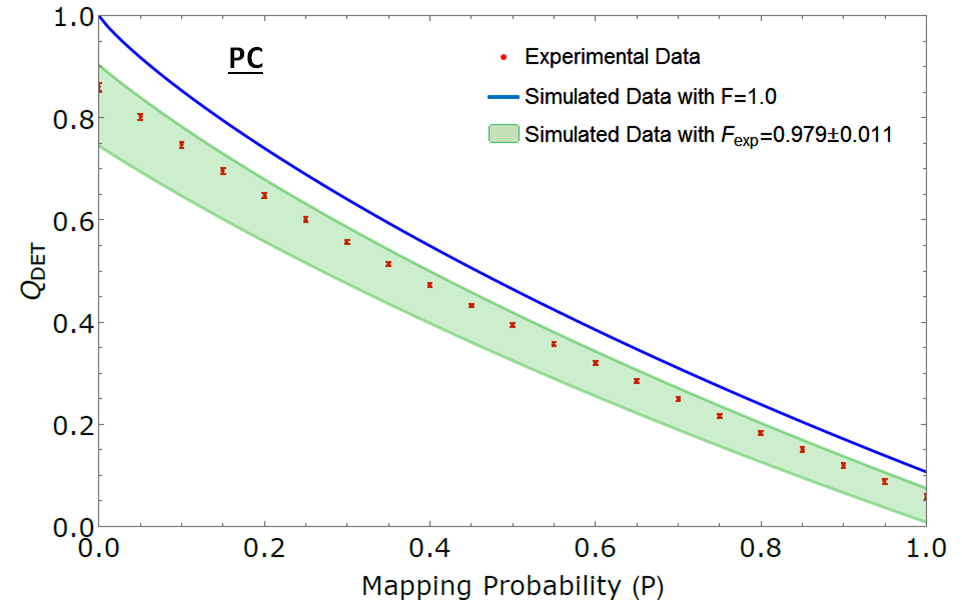}} \caption {
\textbf{Minimal bounds $Q_{DET}$ to the quantum channel capacity.} For
the entire set of data, the red points represent the experimental
values, the continuous blue line corresponds to the ideal simulation
of a pure input state $\ket{\Phi^{+}}$ with fidelity $F=1$. Green shaded areas
correspond to a region of $Q_{DET}$ for an 
input Werner state $\rho_{W}$ within one standard deviation of fidelity $F_{exp}=0.979\pm0.011$. 
\textbf{a) ADC:}
$Q_{DET}$ vs. the damping parameter
$\gamma$. \textbf{b) PDC:} $Q_{DET}$ vs. $p$ for the
statistical mixture of $\mathbb{I}$ and $\sigma_{z}$, with probability vector
$\vec{p}=\{P_{\mathbb{I}},P_{x},P_{y},P_{z}\}=\{1-\frac{p}{2},0,0,\frac{p}{2}\}$. \textbf{c)
DC:} $Q_{DET}$ vs. $p$ for the balanced mixture of $\mathbb{I}$,
$\sigma_{x}$, $\sigma_{y}$ and $\sigma_{z}$, with 
$\vec{p}=\{1-p,\frac{p}{3},\frac{p}{3},\frac{p}{3}\}$. \textbf{d) PC:}
$Q_{DET}$ vs. $p$ for the mixture of 
$\mathbb{I}$, $\sigma_{x}$, $\sigma_{y}$, and $\sigma_{z}$, 
with $\vec{p}=\{1-\frac{p}{6},\frac{p}{12},\frac{p}{18},\frac{p}{36}\}$. The
error bars on the mapping probability are significant only in the case
a), where an uncertainty of $0.5$ degrees in the rotation angle of
$HWP_{V}(\gamma)$ was propagated, while this error was negligible in
the cases b), c) and d). In the case a), the error bars on $Q_{DET}$
were calculated from Poissonian statistics on one set of points,
obtaining negligible values. In the case of b), c) and d), error bars
were obtained from an average of at least 8 sets of points. (\cite{SM} for more details).}
\label{fig:results} 
\end{figure*}

An \textbf{ADC} for polarization qubits with 
probability $\gamma$ can be written as 
${\cal E}(\rho )= K_{0}\rho K_{0}^\dag+K_{1}\rho K_{1}^\dag$, where 
$K_{0}=|H\rangle\langle H|+\sqrt{1-\gamma}|V\rangle\langle V|$ and 
$K_{1}=\sqrt\gamma|H\rangle\langle V|$. 
The experimental setup \cite{amplitude} consists in a Sagnac interferometer (SI) followed by a 
Mach-Zehnder interferometer (MZI), as shown in Fig.\ref{fig:channels}a, that 
allows us to perform the required noise operation. In the Sagnac loop, composed 
by a PBS and three mirrors, a HWP at $0$ degree is put on the path of 
polarization $\ket{H}$, whereas a HWP at $\theta$ is put on the path of 
polarization $\ket{V}$. The MZI recombines the two outputs of the SI in a non-coherent superposition within a Beam Splitter (BS), thus effectively 
performing the damping operation, whose amount $\gamma $ depends on the 
angle $\theta$ \cite{almeida07}.

A \textbf{PDC} for qubits with probability 
$p$ can be written as ${\cal E}(\rho)= \left(1-\frac{p}{2}\right)\rho+\frac{p}
{2} \sigma_z\rho\sigma_z$. This map is achieved by using a sequence of 
two wave retarders: a HWP at 0 degree, which acts as 
$\sigma_z$, and an unrotated Liquid Crystal (LC), which acts as $\mathbb{I}$ or 
$\sigma_{z}$ depending on the applied voltage on the material \cite{liquidC}. The system qubit is sent through the 
combination of this two optical elements, suffering a 
$\sigma_{z}\cdot\sigma_{z}=\mathbb{I}$ or 
$\mathbb{I}\cdot\sigma_{z}=\sigma_{z}$ operations, while the ancilla 
qubit remains untouched. In Fig. 
\ref{fig:channels}b-Top we show the schematic representation of the channel. 
The probabilities $P_{z}=\frac{p}{2}$ and
$P_{\mathbb{I}}=1-P_{z}=1-\frac{p}{2}$ are applied in post-processing, 
by evaluating the ratios of photon coincidences from two separated experiments, $\sigma_{z}$ and
$\mathbb{I}$ respectively. Therefore the total channel corresponds to a mixture of two operations constrained by $P_z + P_{\mathbb {I}} = 1$.

A \textbf{DC} for qubits with probability 
$p$ can be written as ${\cal E}(\rho)
=\left(1-p\right)\rho+\frac{p}{3}\left(\sigma_{x}\rho\sigma_{x}
+\sigma_{y}\rho\sigma_{y}+\sigma_{z}\rho\sigma_{z}\right)$. 
The operations are achieved by the presence (absence) of a rotated HWP 
acting as $\sigma_x$ when it is on the optical path or acting as $\mathbb{I}$ 
if taken off from the path, and by an unrotated LC acting as 
$\mathbb{I}$ or $\sigma_{z}$, depending on the applied voltage.
From the simultaneous actions of the HWP and the LC over 
$\rho$, the operations can be $\mathbb{I}\cdot\mathbb{I}=\mathbb{I}$,
$\sigma_{z}\cdot\mathbb{I}=\sigma_{z}$, 
$\sigma_{z}\cdot\sigma_{x}=i\sigma_{y}$ and 
$\mathbb{I}\cdot\sigma_{x}=\sigma_{x}$.  In Fig. \ref{fig:channels}b-Bottom 
we show the schematic representation of the channel. 
Analogously to the procedure followed for the PDC, 
the effective $\sigma_{z}$, $\sigma_{y}$, $\sigma_{x}$ and $\mathbb{I}$ 
operations were applied in four separated experiments, where 
$P_{z}=P_{y}=P_{x}=\frac{p}{3}$ and $P_{\mathbb{I}}=1-p$ were added in 
post-processing of the experimental outcomes. 

The \textbf{PC} for qubits with probability $p$ can 
be written as ${\cal E}(\rho)=P_{\mathbb{I}}\rho+P_{x}\sigma_{x}\rho
\sigma_{x}+P_{y}\sigma_{y}\rho\sigma_{y}+P_{z}\sigma_{z}\rho\sigma_{z}$ and 
has been implemented by the same procedure used for the DC. 
However, in this case there is no restriction on the 
choice of $P_i$ except for the condition 
$P_{\mathbb{I}}+P_{z}+P_{y}+P_{x}=1$, also valid for \textbf{DC}.\\

\textbf{Results:} In Fig.~\ref{fig:results} we show the experimental measurements of
$Q_{DET}$ for a prepared ADC, PDC, DC and PC. The experimental values are in very good agreement with the theoretical predictions.

These results prove the effectiveness of the method, which assesses
efficiently the lower bound of $Q_{DET}$ to the quantum capacity of noisy
channels with the most common kinds of noise.

Moreover, the well known expressions for the ADC and PDC quantum
capacities \cite{gf,deveshor} coincide with our detectable bound. The
expected non-zero capacity for the ADC occurs when $\gamma<1/2$, while
the certified experimental value was $\gamma<0.45$ (see
Fig. \ref{fig:results}a). For the PDC, we obtained a non vanishing
value of the $Q_{DET}$ for values of $p$ up to $0.65$, whereas the
ideal analytical expression is positive for any value of $p$ (see
Fig. \ref{fig:results}b)).

In the case of DC, the procedure certifies a nonzero channel
capacity for values of $p \leq 0.18$, while the best theoretical lower
bound predicts a nonzero channel capacity up to
$p=0.1892$ \cite{hashing}. Analogously, the technique certifies a nonzero channel
capacity of the chosen PC for any value of $p$, similar to the ideal
simulated channel.\\

\textbf{Conclusions:} We have performed an experiment to detect efficient
lower bounds to the quantum capacity of qubit communication
channels. Our technique does not require any prior knowledge of the
quantum channel, and can be applied to any kind of unknown noise. The
principal feature of the technique resides on the smaller number of
measurement settings with respect to a full process tomography, and is
in very good agreement with the theoretical prediction for the
source we used.  To the best of our knowledge this is the first
experiment where the quantum capacity of a noisy channel is directly
accessed.  Furthermore, the detectable bounds we have provided give
lower bounds to the private information and to the
entanglement-assisted classical capacity, as emphasized in
\cite{ms16}.

These results represent an important step toward an efficient experimental characterization of quantum 
channels such as those used in quantum cryptography, quantum teleportation, and 
quantum dense coding.\\

\textbf{Acknowledgements:} \'A. Cuevas would like to thank the support from the Chilean agency CONICYT and to its PhD scholarships program.

\newpage

\widetext

\appendix*
\section{Supplemental Material}

\begin{center}
\textbf{On the effect of noise of the input state in the experimental 
detection of quantum channel capacities.}
\end{center}

The theoretical results of Ref. \cite{ms16}, summarized in Eqs. (3--5) of the 
Letter, were derived under the assumption of sending a pure bipartite state 
at the input of the unknown channel. Realistically, as in the present 
experimental set-up, the generated input state will be affected by noise, 
thus producing a {\em mixed} state at the channel input.
 
Specifically, since our source indeed generates a maximally entangled
state affected by isotropic noise, in the following, for arbitrary
finite dimension $d$, we consider an isotropic noise map $\cal N$,
leading to a bipartite mixed input state $\nu $ given by the convex
combination of a maximally entangled state $|\Phi ^+ \rangle =\sum
_{n=0}^{d-1}|n\rangle |n \rangle /\sqrt d$ and the totally mixed state
$\frac{1}{d^2}I\otimes I$.  Then, we can write 
\begin{eqnarray} \nu
\equiv (I \otimes {\cal N})\ket{\Phi ^+}\bra{\Phi ^+} = ({\cal
N}\otimes I) \ket{\Phi ^+}\bra{\Phi ^+} = \frac{d^2 F -1}{d^2 -1}
\ket{\Phi ^+}\bra{\Phi ^+} + \frac {1-F}{d^2 -1}I\otimes I
\;,\label{iso}\end{eqnarray} 
in terms of the fidelity F with the ideal
maximally entangled state, namely $F=\langle \Phi ^+ | \nu |\Phi^+
\rangle $.

In the case of $d=2$ we have
$\nu\rightarrow\rho_{W}\equiv\frac{4F-1}{3}
\ket{\Phi^{+}}\bra{\Phi^{+}}+\frac{1-F}{3}I\otimes I$, also known as
Werner state.
In our work, the experimental input density matrix $\nu$ is calculated
by a Mathematica algorithm based on two steps. In the first one, a
preliminary matrix $\nu'$ is computed by considering the registered
photon coincidences of 36 suitable plates rotations as projections of
the experimental state. In the second step, the algorithm imposes
positivity and forces $\nu '$ to approach a legitimate positive
matrix $\nu$.  This is done by a Maximum Likelihood Maximization
process, based on Poisson statistics, resulting in a physical version
of $\nu$, which is now a positive
matrix with unit trace \cite{jkmw01}. 
Then, the fidelity $F$ is obtained by the straightforward evaluation of 
$\langle \Phi ^+ | \nu |\Phi^+ \rangle $.

As long as the noise map $\cal N$ acts on maximally entangled states,
its action can be equivalently ascribed to the system or the ancilla
qudit. For this reason the use of a noisy bipartite state as in
Eq. (\ref{iso}) for our detection protocol is equivalent to having a
perfect input $|\Phi ^+\rangle $ along with a quantum measurement
degraded by the isotropic dual map ${\cal N}^\vee$ (the map in the
Heisenberg picture). In fact, for any measurement basis $\{|\Phi _i
\rangle \}$ the reconstructed probabilities are \begin{eqnarray} p_i
&&= \Tr [({\cal I}_A \otimes {\cal E}) ({\cal I}_A \otimes {\cal N})
(|\Phi ^+ \rangle \langle \Phi ^+ |) |\Phi _i \rangle\langle\Phi _i|]
=\Tr [({\cal I}_A \otimes {\cal E}) (|\Phi ^+ \rangle \langle \Phi ^+
|) ({\cal N}^\vee \otimes {\cal I} ) (|\Phi _i \rangle\langle\Phi
_i|)] \nonumber \\& & =\Tr [({\cal I}_A \otimes {\cal E}) (|\Phi ^+
\rangle \langle \Phi ^+ |) \Pi _i]\;, \label{pimeas3} \end{eqnarray}
where \begin{eqnarray} \Pi _i \equiv ({\cal N}^\vee \otimes {\cal I} )
(|\Phi _i \rangle\langle\Phi _i|) \;\label{pom} \end{eqnarray} is
generally the element of a POVM.  Then, in order to take into account
the effect of noise, we need to generalize the bound of
Ref. \cite{ms16} \begin{eqnarray} S_e\left (\rho , {\cal E} \right
)\leq H (\vec p)\;, \label{se-bound} \end{eqnarray} to the case of a
probability vector $\vec p$ pertaining to an arbitrary POVM $\{ \Pi _i
\}$ for the tensor product of the reference and system Hilbert spaces,
where \begin{eqnarray} p_i = \Tr [({\cal I}_A \otimes {\cal E})(|\Psi
_\rho \rangle \langle \Psi _\rho |) \Pi _i ] \;.  \label{pimeas2}
\end{eqnarray} In the following we provide such a generalization. Let
us consider a density matrix $\sigma$, along with its spectral
decomposition $\sigma =\sum _j s_j |\phi _j \rangle \langle \phi _j
|$, and let us define the conditional probability $p(i|j)=\langle
\phi_j | \Pi _i | \phi _j \rangle $. Clearly, one has $p_i \equiv \Tr
[\sigma \Pi _i] = \sum _j s_j p(i|j)$. Then, \begin{eqnarray} S(\sigma
)- H(\vec p) &=& \sum _i p_i \log _2 p_i - \sum _j s_j \log _2 s_j =
\sum _{i,j} s_j p(i|j)( \log _2 p_i -\log _2 s_j ) \nonumber \\ & \leq
& \log_2 \left (\sum_{i,j}s_j p(i|j) \frac{p_i}{s_j} \right ) = \log_2
\vec r \cdot \vec p \;, \end{eqnarray} where we used Jensen's
inequality, and defined $\vec r$ with vector components $r_i = \sum _j
\bra{\phi _ j} \Pi _i \ket {\phi _j }$.  Upon choosing $\sigma =
({\cal I}_R \otimes {\cal E})(|\Psi _\rho \rangle \langle \Psi _\rho
|) $, one obtains the more general bound 
\begin{eqnarray} S_e\left
(\rho , {\cal E} \right )\equiv S(\sigma ) \leq H (\vec p)+ \log _2
\vec t \cdot \vec p\;, \label{se-bound2} 
\end{eqnarray} 
with $p_i $ as
in Eq. (\ref{pimeas2}) and $t_i \equiv \Tr [\Pi _i] \geq r_i$.  Then,
the detected quantum capacity $Q_{DET}$ in Eq. (4) of the Letter is
simply replaced with 
\begin{eqnarray} Q_{DET}=S\left [{\cal E} (\rho
)\right ] -H(\vec p) - \log _2 \vec t \cdot \vec p \;, \label{qnew}
\end{eqnarray} 
in terms of the noisy reconstructed probabilities of
Eq. (\ref{pimeas3}).

Notice now that for any unital noise map (as the isotropic-noise one), the dual map is trace-preserving. Hence, if $\Pi_i $ is of the form as in Eq. (\ref{pom}), one has $t_i \equiv \Tr [\Pi _i]=1$. Then, 
the last term in (\ref{qnew}) vanishes, namely $\log _2 \vec t \cdot \vec p =0$.

Let us see the effect of noise on the quantum capacity detection for specific 
channels.\\

\noindent 1) Amplitude Damping Channel (ADC) for qubits
\begin{eqnarray}
{\cal E}(\rho )= K_0 \rho K_0^\dag + K_1  \rho K_1^\dag \;, 
\end{eqnarray}
where 
$K_0= |0 \rangle \langle 0| + \sqrt {1- \gamma }|1 \rangle \langle 1|$ and 
$K_1= \sqrt \gamma |0 \rangle \langle 1|$. 

For an input state as in Eq. (\ref{iso}), the bipartite output is given by 
\begin{eqnarray}
({\cal I}_A\otimes {\cal E}) \nu  &=&
\left ( c_1 \gamma _+ + c_2 \right ) 
|\Phi ^+ \rangle \langle \Phi ^+|  
+ \left ( c_1 \gamma _- + c_2  \right ) 
|\Phi ^- \rangle \langle \Phi ^-|  
+\frac \gamma 4 
(|\Phi ^+ \rangle \langle \Phi ^-|+|\Phi ^- \rangle \langle \Phi ^+|)
\nonumber \\& + &  \frac 1 4 (\gamma c_1 + 4c_2 )
(|\Psi ^+ \rangle \langle \Psi ^+| + |\Psi ^- \rangle \langle \Psi ^-|)
- \frac \gamma 4 (|\Psi ^+ \rangle \langle \Psi ^-| + |\Psi ^- \rangle \langle \Psi ^+|)
\;,\label{outdamp} 
\end{eqnarray}
with $\gamma _\pm =\frac 14 (1\pm \sqrt{1- \gamma })^2$, 
$c_1=(4F-1)/3$, and $c_2=(1-F)/3$.  
The reduced output state is given by ${\cal E}\left (\frac I2\right )= 
\frac 12 (I +\gamma \sigma _z)$, 
hence  it has von Neumann entropy 
$S\left [{\cal E}\left (\frac I 2 \right )\right ]
=H_2 \left(\frac {1- \gamma }{2}\right )$. 
By performing the local measurement of $ \sigma _x 
\otimes \sigma _x $, $ \sigma _y \otimes \sigma _y $, and $ \sigma _z 
\otimes \sigma _z $, estimating the von Neumann entropy 
$S\left [{\cal E}\left (\frac I 2 \right )\right ]$, 
and optimising $\vec p$, one can detect the bound 
\begin{eqnarray}
Q\geq Q_{DET}&=& H_2 \left(\frac {1- \gamma }{2}\right )- H(\vec p)  
\,,\label{bounddamp}
\end{eqnarray}
where the optimal vector of probabilities is given by
\begin{eqnarray}
 \vec{p}&=& \left ( \frac{
2 + 4 F (1 - \gamma ) + \gamma  + \sqrt{
 4 (1 - 4 F)^2 (1-\gamma ) + 9 \gamma ^2} }{12}, 
\frac{ 2 + 4 F (1 - \gamma ) + \gamma  - \sqrt{
 4 (1 - 4 F)^2 (1-\gamma ) + 9 \gamma ^2}}{12}, \right. 
\nonumber \\& & \left.
\frac {(1-F)(1-\gamma )}{3} , \frac {2 +\gamma -2F(1-\gamma )}{6} \right)
\;. 
\end{eqnarray}
Such probabilities correspond to the optimal basis
\begin{eqnarray}
\{ a |\Phi ^+ \rangle + b |\Phi ^- \rangle  , 
-b |\Phi ^+ \rangle + a |\Phi ^- \rangle  ,  
\frac{1}{\sqrt 2}(|\Psi ^+ \rangle +  |\Psi ^- \rangle )\equiv |01 \rangle   , 
\frac{1}{\sqrt 2}(|\Psi ^+ \rangle -  |\Psi ^- \rangle )\equiv |10 \rangle    
\}\;,
\end{eqnarray}
where  
\begin{eqnarray}
a=\sqrt{\frac{\cosh \eta + 1}{2\cosh \eta }}\,, \qquad 
b=\sqrt{\frac{\cosh \eta - 1}{2\cosh \eta }}\,, 
\end{eqnarray}
with $\eta\equiv\text{arcsinh}\left (\frac{3\gamma}{2(4F-1)\sqrt{1-\gamma}} \right )$.  

The case of amplitude damping channel is a relevant example showing that the Bell basis can be sub-optimal 
for the quantum capacity certification.\\

\noindent 2) For the Phase Damping Channel (PDC) for qubits
\begin{eqnarray}
{\cal E}(\rho )= \left (1-\frac p 2 \right ) 
\rho + \frac p 2 \sigma _z \rho \sigma _z
\,, 
\end{eqnarray}
one has   
\begin{eqnarray}
Q=Q_1= 1 - H_2 \left( \frac p2 \right ) 
\geq Q_{DET}=1- H(\vec p)\,,
\end{eqnarray}
where the vector of probabilities $\vec p$ is given by 
\begin{eqnarray}
\vec p= 
\left\{ \left (1- \frac p2 \right ) F + \frac p2  \frac{1-F}{3} 
,   \frac p2  F + \left (1 -\frac p2 \right ) \frac{1-F}{3}, 
\frac{1-F}{3},  \frac{1-F}{3} \right \}\;,
\end{eqnarray}
which corresponds to the Bell basis.\\

\noindent 3) For the Depolarizing Channel (DC) in dimension $d$
\begin{eqnarray}
{\cal E}(\rho )=
\left ( 1 - p \frac {d^2}{d^2-1}  \right )\rho  + p 
\frac {d^2}{d^2 -1}  \frac Id \;, 
\end{eqnarray}
the detectable bound is now  
\begin{eqnarray}
Q \geq Q_{DET}=\log _2 d - H_2 (p') - p'\log _2 (d^2 -1)\;,\label{hashd}
\end{eqnarray}
where 
\begin{eqnarray}
p'=\frac { d^2 [1-F(1-p)]+ F-p-1] }{d^2 -1}
\;.
\end{eqnarray}
For qubits
\begin{eqnarray}
Q \geq Q_{DET}=1 - H_2 (p') - p'\log _2 3\;,\label{hashd2}
\end{eqnarray}
with $p'=(1-F)+ \frac p3 (4F -1)$. The reconstructed probabilities for the Bell basis $\{|\Phi ^+ \rangle , |\Phi ^- \rangle , |\Psi ^+ \rangle , |\Psi ^- \rangle \}$ correspond to $\{1-p',\frac{p'}{3},\frac{p'}{3},\frac{p'}{3}\}$.\\

\noindent 4) For a Pauli Channel (PC) in dimension $d$
\begin{eqnarray}
 {\cal E}(\rho )=\sum _{m,n=0}^{d-1} p_{mn} U_{mn} \rho U^{\dag }_{mn}\;,
\end{eqnarray}
with $U_{mn}=\sum _{k=0}^{d-1} e^{\frac{2\pi i}{d} km} |k \rangle 
\langle (k + n)\!\!\!\mod d |$, one has 
\begin{eqnarray}
Q \geq Q_{DET}=\log _2 d - H({\vec p}\,') \;, 
\end{eqnarray}
where ${\vec p}\, '$ is the $d^2$-dimensional vector of probabilities 
pertaining to the generalised Bell projectors, whose components are given by 
\begin{eqnarray}
p_{mn}'= \frac{1}{d^2-1}[(d^2F -1)p_{mn}+1-F]
\;.
\end{eqnarray}
For the qubit case,  ${\cal E}(\rho )=\sum _{i=0}^3 p_i \sigma _i \rho \sigma _i $, 
and $p_i'=\frac 13[(4F-1)p_i +1-F]$. 
\newpage

\begin{center}
\textbf{Experimental Expectation Values}
\end{center}

The expanded expressions of the probability vectors $p_{i,j}=\braket{\Pi_{i,j}}$ used in our experiment obey the following form, described here for $B_{1}$:
\begin{align}
p_{1,1}&=\braket{(a\ket{\Phi^{+}}+b\ket{\Phi^{-}})(a\bra{\Phi^{+}}+b\bra{\Phi^{-}})}\nonumber\\
&=\frac{1}{4}(\braket{\mathbb{I}\otimes\mathbb{I}}+\braket{\sigma_{z}\otimes\sigma_{z}})+\frac{a b}{2}(\braket{\sigma_{z}\otimes\mathbb{I}}+\braket{\mathbb{I}\otimes\sigma_{z}})+\frac{a^{2}-b^{2}}{4}(\braket{\sigma_{x}\otimes\sigma_{x}}-\braket{\sigma_{y}\otimes\sigma_{y}})\;, 
\\  
p_{1,2}&=\braket{(-b\ket{\Phi^{+}}+a\ket{\Phi^{-}})(-b\bra{\Phi^{+}}+a\bra{\Phi^{-}})}\nonumber\\
&=\frac{1}{4}(\braket{\mathbb{I}\otimes\mathbb{I}}+\braket{\sigma_{z}\otimes\sigma_{z}})-\frac{a b}{2}(\braket{\sigma_{z}\otimes\mathbb{I}}+\braket{\mathbb{I}\otimes\sigma_{z}})-\frac{a^{2}-b^{2}}{4}(\braket{\sigma_{x}\otimes\sigma_{x}}-\braket{\sigma_{y}\otimes\sigma_{y}}) \;,
\\  
p_{1,3}&=\braket{(c\ket{\Psi^{+}}+d\ket{\Psi^{-}})(c\bra{\Psi^{+}}+d\bra{\Psi^{-}})}\nonumber\\
&=\frac{1}{4}(\braket{\mathbb{I}\otimes\mathbb{I}}-\braket{\sigma_{z}\otimes\sigma_{z}})+\frac{cd}{2}(\braket{\sigma_{z}\otimes\mathbb{I}}-\braket{\mathbb{I}\otimes\sigma_{z}})+\frac{c^{2}-d^{2}}{4}(\braket{\sigma_{x}\otimes\sigma_{x}}+\braket{\sigma_{y}\otimes\sigma_{y}}) \;, \\
p_{1,4}&=\braket{(-d\ket{\Psi^{+}}+c\ket{\Psi^{-}})(-d\bra{\Psi^{+}}+c\bra{\Psi^{-}})}\nonumber\\
&=\frac{1}{4}(\braket{\mathbb{I}\otimes\mathbb{I}}-\braket{\sigma_{z}\otimes\sigma_{z}})-\frac{cd}{2}(\braket{\sigma_{z}\otimes\mathbb{I}}-\braket{\mathbb{I}\otimes\sigma_{z}})-\frac{c^{2}-d^{2}}{4}(\braket{\sigma_{x}\otimes\sigma_{x}}+\braket{\sigma_{y}\otimes\sigma_{y}})\;
\end{align}
The rest of the probability vectors, $\vec{p}_{2,j}$ and $\vec{p}_{3,j}$, can be calculated analogously.

Our protocol only needs the following expectation values of observables on the joint ancilla-system state
\begin{align}
&\braket{\sigma_{z}\otimes\sigma_{z}}=
\frac{1}{N_{z}}CC(\ket{H}\bra{H} \otimes\ket{H}\bra{H} -\ket{H}\bra{H} \otimes\ket{V}\bra{V} 
-\ket{V}\bra{V} \otimes\ket{H}\bra{H} +\ket{V}\bra{V} \otimes\ket{V}\bra{V} ) 
,\\
&\braket{\sigma_{y}\otimes\sigma_{y}}=
\frac{1}{N_{y}}CC(\ket{R}\bra{R} \otimes\ket{R}\bra{R} -\ket{R}\bra{R} \otimes\ket{L}\bra{L} 
-\ket{L}\bra{L} \otimes\ket{R}\bra{R} +\ket{L}\bra{L} \otimes\ket{L}\bra{L} ) 
,\\
&\braket{\sigma_{x}\otimes\sigma_{x}}= 
\frac{1}{N_{x}}CC(\ket{+}\bra{+} \otimes\ket{+}\bra{+} -\ket{+}\bra{+} \otimes\ket{-}\bra{-} 
-\ket{-}\bra{-} \otimes\ket{+}\bra{+} +\ket{-}\bra{-} \otimes\ket{-}\bra{-} ) 
,\\
& \braket{\mathbb{I}\otimes\mathbb{I}}=\frac{1}{N_{z}}CC(\ket{H}\bra{H} \otimes\ket{H}\bra{H} +\ket{H}\bra{H} \otimes\ket{V}\bra{V} 
+\ket{V}\bra{V} \otimes\ket{H}\bra{H} +\ket{V}\bra{V} \otimes\ket{V}\bra{V} )
,\\
&\braket{\sigma_{z}\otimes\mathbb{I}} = 
\frac{1}{N_{z}}CC(\ket{H}\bra{H} \otimes\ket{H}\bra{H} +\ket{H}\bra{H} \otimes\ket{V}\bra{V} 
-\ket{V}\bra{V} \otimes\ket{H}\bra{H} -\ket{V}\bra{V} \otimes\ket{V}\bra{V} ) \label{ZI}
,\\
&\braket{\sigma_{y}\otimes\mathbb{I}}=
\frac{1}{N_{y}}CC(\ket{R}\bra{R} \otimes\ket{R}\bra{R} +\ket{R}\bra{R} \otimes\ket{L}\bra{L} 
-\ket{L}\bra{L} \otimes\ket{R}\bra{R} -\ket{L}\bra{L} \otimes\ket{L}\bra{L} ) 
\label{YI}
,\\
&\braket{\sigma_{x}\otimes\mathbb{I}}=  
\frac{1}{N_{x}} CC(\ket{+}\bra{+} \otimes\ket{+}\bra{+} +\ket{+}\bra{+} \otimes\ket{-}\bra{-} 
-\ket{-}\bra{-} \otimes\ket{+}\bra{+} -\ket{-}\bra{-} \otimes\ket{-}\bra{-} ) 
\label{XI} 
,\\
&\braket{\mathbb{I}\otimes\sigma_{z}}=  
\frac{1}{N_{z}}CC(\ket{H}\bra{H} \otimes\ket{H}\bra{H} -\ket{H}\bra{H} \otimes\ket{V}\bra{V} 
+\ket{V}\bra{V} \otimes\ket{H}\bra{H} -\ket{V}\bra{V} \otimes\ket{V}\bra{V} ) \label{IZ} 
,\\
&\braket{\mathbb{I}\otimes\sigma_{y}}=  
\frac{1}{N_{y}}CC(\ket{R}\bra{R} \otimes\ket{R}\bra{R} -\ket{R}\bra{R} \otimes\ket{L}\bra{L} 
+\ket{L}\bra{L} \otimes\ket{R}\bra{R} -\ket{L}\bra{L} \otimes\ket{L}\bra{L} ) 
\label{IY} 
,\\
&\braket{\mathbb{I}\otimes\sigma_{x}}= 
\frac{1}{N_{x}}CC(\ket{+}\bra{+} \otimes\ket{+}\bra{+} -\ket{+}\bra{+} \otimes\ket{-}\bra{-} 
+\ket{-}\bra{-} \otimes\ket{+}\bra{+} -\ket{-}\bra{-} \otimes\ket{-}\bra{-} ), 
\label{IX} 
\end{align}
with
\begin{align}
&N_{z}=CC(\ket{H}\bra{H} \otimes\ket{H}\bra{H} +\ket{H}\bra{H} \otimes\ket{V}\bra{V} 
+\ket{V}\bra{V} \otimes\ket{H}\bra{H} +\ket{V}\bra{V} \otimes\ket{V}\bra{V} )
,\\
&N_{y}=CC(\ket{R}\bra{R} \otimes\ket{R}\bra{R} +\ket{R}\bra{R} \otimes\ket{L}\bra{L} 
+\ket{L}\bra{L} \otimes\ket{R}\bra{R} +\ket{L}\bra{L} \otimes\ket{L}\bra{L} )
,\\
&N_{x}=CC(\ket{+}\bra{+} \otimes\ket{+}\bra{+} +\ket{+}\bra{+} \otimes\ket{-}\bra{-} 
+\ket{-}\bra{-} \otimes\ket{+}\bra{+} +\ket{-}\bra{-} \otimes\ket{-}\bra{-} ) .
\end{align}
In the above equations  
$CC(\ket{i}\bra{i} \otimes \ket{j}\bra{j})$
denotes  the coincident photon 
detections of the state within a time window, associated to the 
projection on the state  $\ket{i}\otimes \ket{j}$ of the ancilla and system qubits, with 
$i,j=H,V,L,R,+,-$. 
The different bases are characterized as logical $\{\ket{H},\ket{V}\}$, circular 
$\{\ket{L}=\frac{1}{\sqrt{2}}(\ket{H}+i\ket{V}),\ket{R}=\frac{1}{\sqrt{2}}(\ket{H}-i\ket{V}) \}$, or diagonal 
$\{\ket{+}=\frac{1}{\sqrt{2}}(\ket{H}+\ket{V}),\ket{-}=\frac{1}{\sqrt{2}}(\ket{H}-\ket{V})\}$, while $N_{z}$, $N_{y}$,  
and $N_{x}$ are the normalization factors also expressed in terms of 
coincidences.

According to the above equations, all the expectation values are obtained just 
by the measurement of 12 polarization projections of the state (4 by each of the 3 observables), 
which makes the process efficient in terms registration and analysis of data, given the reduced number of operations compared with a standard process tomography. 
Equations (\ref{ZI}-\ref{XI}) are equivalent to take a partial trace of the 
S-qubit, and measure exclusively the A-qubit by triggering the 
coincidences with the detection of S. Similarly, equations (\ref{IZ}-\ref{IX}) 
are equivalent to take a partial trace of the A-qubit, and measure 
exclusively the S-qubit by triggering the coincidences with the detection 
of A.\\

\begin{center}
\textbf{Photon Count Rates and Detection Efficiencies}
\end{center}

The overall detection efficiency of single photons in the system or ancilla modes of our experiment is given approximately by
\begin{equation}
\epsilon_{overall}\approx\epsilon_{source}\cdot\epsilon_{exp}\cdot\epsilon_{tomo}\cdot\epsilon_{SMF}\cdot\epsilon_{APD}
\end{equation}

Where $\epsilon_{source}\approx95\%$ is the transmissivity of down converted photons through the optical elements of the quantum source, $\epsilon_{exp}$ is the transmissivity of the tested channel, $\epsilon_{tomo}\approx99\%$ is the transmissivity of the tomography optics, $\epsilon_{SMF}\approx73\%$ is the coupling efficiency of photons within a single mode optic fiber (SMF) and $\epsilon_{APD}\approx70\%$ is the quantum efficiency of the Avalanche Photodetectors (APDs). 

The system-S photons of the source were transmitted to the tested channels through a polarization compensated SMF link. Then, the effective efficiency of any kind of channel was composed by 

\begin{equation}
\epsilon_{exp}\approx\epsilon_{SMF}\cdot\epsilon_{channel}
\end{equation}

Where $\epsilon_{channel}$ corresponds to transmissivity of each bulk channel, ranging from $\approx98\%$ in the case of PDC, DC and PC, to $\approx60\%$ for the ADC. Thus, without testing any kind of channel inside the mode ($\epsilon_{exp}=1$) the associated overall detection efficiency for single photons was $\epsilon_{overall}\approx48\%$. In the case of PDC, DC or PC the overall detection efficiency was $\approx34\%$, and for ADC was $\approx21\%$.

Considering $2.5$mW of laser pumping power and the overall single photon efficiency $\epsilon_{overall}\approx48\%$ from the SPDC process to the single photon detection in two synchronized APDs within a time window of 6ns, we get an effective source generation of nearly $C_{S}=C_{A}=375000\frac{\text{single photons}}{\text{sec}}$ in the S-A modes and $C_{SA}=60000\frac{\text{coincident photons}}{\text{sec}}$ without the presence of any channel on S. These values corresponds to a heralding efficiency of 
\begin{equation}
\eta=\frac{C_{SA}}{\sqrt{C_{S}\cdot C_{A}}}\approx16\%
\end{equation}

From the above quantities it is also possible to calculate the real number of photons created by the source, right after the SPDC generation, even if we don't have access to them. To estimate that quantity we assume that $C_{S}=C_{A}$, then the source output counts will be
\begin{eqnarray}
C_{S,Source}&=&\frac{C_{S}}{\epsilon_{overall}}\approx 781000\frac{\text{single photons}}{sec}\\
C_{SA,Source}&=&\frac{C_{SA}}{(\epsilon_{overall})^{2}}\approx 260000\frac{\text{coincident photons}}{sec}
\end{eqnarray}

The registered dark counts were $<500\frac{\text{single photons}}{sec}$, and an accidental coincidence rate $<5\%$ was evaluated.

\begin{center}
\textbf{Experimental Error Analysis}
\end{center}

Each experimental error bar seen in our results originates from three main contributions:
\begin{enumerate}
\item	Poissonian statistics on the coincidence photons counts, which is propagated using the a Monte Carlo simulation around the mean coincidence values registered in 5 seconds of integration.   
\item	The experimental channels can have systematic errors due an imperfect balance between the internal maps, or also random errors as 
the propagation of the uncertainty in the rotation of the HWP and LC.
\item	Each experimental point is a mean of a different number of repetitions of the same experiment, so that we considered the standard deviation as a contribution to the random error. 
\end{enumerate}

\begin{itemize}
\item \textbf{For the ADC:} The principal contribution to the total error originates from the uncertainty of the dumping preparation, namely the rotation of the $HWP_{V}(\varphi)=\left(\begin{matrix}
cos(2\varphi)&-sin(2\varphi)\\
-sin(2\varphi)&-cos(2\varphi)\end{matrix}\right)$. In this case, there is a direct connection of the damping to the angle $\varphi$ of the plate through the equation
\begin{equation}
\varphi=\frac{arccos(-\sqrt{1-\gamma})}{2}
\end{equation}
Given the non-linearity of dumping, an uncertainty of $\Delta=0.5[\text{degrees}]$ can be very sensitive to any change.

Any other source of error, including the Poissonian distribution of the counts and its proper propagation is practically negligible compared with the propagation of $\Delta$. Thus the presentation of the $Q_{Det}$ under the action of an ADC only have effective error bars in the x-axis.

\item \textbf{For the PC:} Unlike the ADC, the PC is achieved by the combination of four different experiments ($\mathbb{I},\sigma_{x},\sigma_{y},\sigma_{z}$), which are prepared separately as a combinations of one HWP and one LC. 

Even if we have more than one optical element with uncertainty $\Delta$, its error propagation represents a very small contribution to the total error. For example, let's consider a more general channel $\Lambda(\rho)=\sum_{i=0}^{3}p_{i}A_{i}\rho A_{i}^{\dagger}$ with $p_{0}+p_{1}+p_{2}+p_{3}=1$, and where each $A_{i}$ map is composed by two LCs, which have an effective operation of $\Gamma_{low}(\varphi)=\left(\begin{matrix}
cos(2\varphi)&-sin(2\varphi)\\
-sin(2\varphi)&-cos(2\varphi)\end{matrix}\right)$ for a low tension value, and $\Gamma_{high}=\mathbb{I}$ for high tension values.

If we want to reproduce the imprecise Pauli channel, namely with an uncertainty of $\Delta$ for every rotation dependence of the LC's, we get that
\begin{eqnarray}
\mathbb{I}=\sigma_{z}\circ\sigma_{z}&\text{      is achieved by      }&A_{0}=\Gamma_{low}(0\pm\Delta)\circ\Gamma_{low}(0\pm\Delta)\\
\sigma_{x}=\sigma_{x}\circ\mathbb{I}&\text{      is achieved by      }&A_{1}=\Gamma_{low}(\pi/2\pm\Delta)\circ\Gamma_{high}\\
\sigma_{y}=i\sigma_{x}\circ\sigma_{z}&\text{      is achieved by      }&A_{2}=\Gamma_{low}(\pi/2\pm\Delta)\circ\Gamma_{low}(0\pm\Delta)\\
\sigma_{z}=\mathbb{I}\circ\sigma_{z}&\text{      is achieved by      }&A_{3}=\Gamma_{high}\circ\Gamma_{low}(0\pm\Delta)
\end{eqnarray}

It is not difficult to verify that $\Lambda$ transforms the state $\rho$ almost in the same way for any propagation of errors within $\Delta$. Then, $\Lambda$ applies the same degree of decoherence to $\rho$ for a propagated error within $\Delta$. The direct consequence of this fact is a negligible error contribution from the angle uncertainty to the final quantum channel capacity.

In this kind of channels the principal error contribution originates from the combination of state measurements obtained in different experiments, because of unavoidable photon counts fluctuations. Any other source of error, including the Poissonian distribution of the counts and its proper propagation is negligible. Thus, the presentation of the $Q_{DET}$ under the action of a PC (and also for \textbf{DC and PDC}) only have effective error bars in the y-axis. 

\end{itemize}


\begin{thebibliography}{99}
\bibitem{nota} This can also be seen by the following identity for the expectations 
on the bipartite output state $\langle\sigma_\alpha\otimes\sigma_\beta\rangle =
\Tr[({\cal I}_A\otimes{\cal E})(|\Phi^+\rangle\langle\Phi^+|)(\sigma_\alpha\otimes\sigma_\beta)]=
\frac{1}{2}\Tr[\sigma_\beta {\cal E}(\sigma_\alpha^{\tau})]$, where $\tau $ denotes the transposition. The expectation values $\langle\sigma_\alpha\otimes\sigma_\alpha\rangle$ and therefore also the probabilities in Eq. (\ref{pimeas}) can then be obtained by considering only the system qubit, preparing it in the eigenstates of $\sigma_\alpha^\tau$ with equal probabilities, and measuring $\sigma_\alpha $ at the output of the channel.
\end{thebibliography}
\end{document}